\documentclass[onecolumn,showpacs,pre,floatfix,amsmath,amssymb,notitlepage,superscriptaddress]{revtex4-1}

\usepackage{bm,bbm}
\usepackage{graphicx,epsfig,subfigure}
\usepackage{amsmath,amssymb}
\usepackage{amsthm}
\usepackage{color,xcolor}
\usepackage{url}
\usepackage[utf8]{inputenc}
\usepackage{enumitem}
\usepackage{setspace}
\usepackage{dcolumn}
\usepackage{bm}
\usepackage{times}
\usepackage{enumerate}
\usepackage{footnote}
\usepackage[titletoc]{appendix}
\usepackage{lineno}
\usepackage{algorithmic} 
\usepackage{algorithm} 

\usepackage{mathbbol}
\usepackage{bm}
\usepackage{bbm}
\usepackage{graphicx}
\usepackage{url}
\usepackage{xcolor}
\usepackage{enumitem}

\usepackage{hyperref}
\hypersetup{
    colorlinks=true,
    linkcolor=blue,
    filecolor=magenta,      
    urlcolor=cyan,
    pdftitle={Overleaf Example},
    pdfpagemode=FullScreen,
    }

\newlength\myindent
\newcommand\bindent{%
	\begingroup
	\setlength{\itemindent}{\myindent}
	\addtolength{\algorithmicindent}{\myindent}
}

\newcommand\eindent{\endgroup}

\begin{document}

\title{Skeleton coupling: a novel interlayer mapping of community evolution in temporal networks}
\author{Bengier Ülgen Kilic}\email{bengieru@buffalo.edu}
\affiliation{Department of Mathematics, University at Buffalo, SUNY, New York, USA}
\author{Sarah Feldt Muldoon}\email{smuldoon@buffalo.edu}
\affiliation{Department of Mathematics, University at Buffalo, SUNY, New York, USA}
\affiliation{CDSE program, University at Buffalo, SUNY, New York, USA}
\affiliation{Neuroscience program, University at Buffalo, SUNY, New York, USA}

\date{\today} 

\clearpage

\begin{abstract} 
\begin{center}
    \textbf{Abstract}
\end{center}
Dynamic community detection (DCD) in temporal networks is a complicated task that involves the selection of a method and its associated hyperparameters.  How to choose the most appropriate method generally depends on the type of network being analyzed and the specific properties of the data that define the network.  In functional temporal networks derived from neuronal spike train data, communities are expected to be transient, and it is common for the network to contain multiple singleton communities.  Here, we compare the performance of different DCD methods on functional temporal networks built from synthetic neuronal time series data with known community structure.  We find that, for these networks, DCD methods that utilize interlayer links to perform community carryover between layers outperform other methods.  However, we also observe that DCD performance is highly dependent on the topology of interlayer links, especially in the presence of singleton and transient communities.  We therefore define a novel way of defining interlayer links in temporal networks called skeleton coupling that is specifically designed to enhance the linkage of communities in the network throughout time based on the topological properties of the community history.  We show that integrating skeleton coupling with current DCD methods improves the method's performance in synthetic data with planted singleton and transient communities.  The use of skeleton coupling to perform DCD will therefore allow for more accurate and interpretable results of community evolution in real-world neuronal data or in other systems with transient structure and singleton communities.

\end{abstract}

\maketitle

\clearpage

\section{Introduction}

Complex systems are often composed of elements whose dynamics and interactions can change over time.  Such temporal events might describe human communication \cite{comnet1}, proximity \cite{socialnet1,socialnet2,socialnet3}, trade and transportation \cite{tradenet1,transpnet1},  citation and collaboration \cite{scicollab1,scicollab2}, or biological \cite{bionet1,bionet2} and neuronal interactions \cite{neuronalnet1,neuronalnet2}. Modeling these systems as temporal networks \cite{tempnet1,tempnet2} can be useful, as network nodes and edges can capture temporal properties of the data.  This is particularly relevant for systems with nodes whose dynamics can be represented using time series data. 

\begin{figure}[!h]
	\centering
	\includegraphics[width = 1\linewidth]{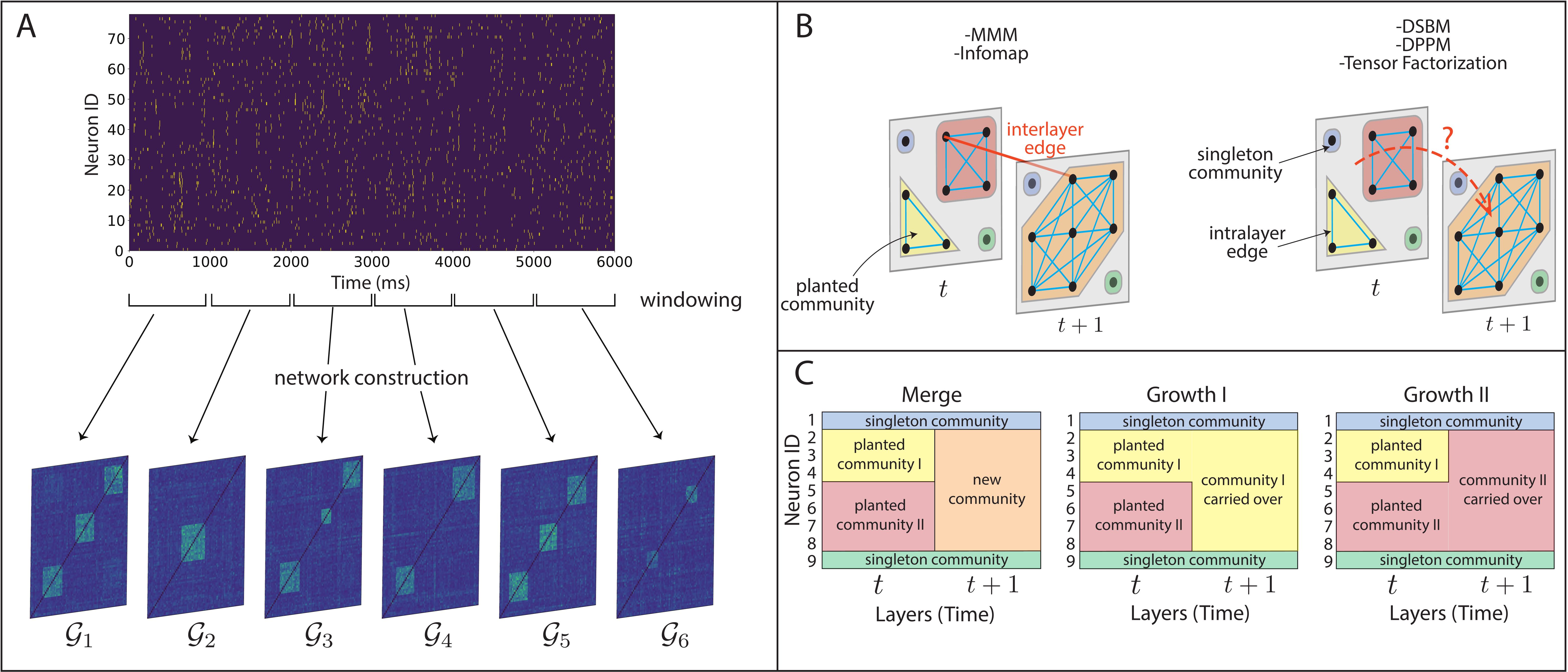}
	\caption{
		\textbf{From time-series to dynamic community analysis}
		\textbf{A.} A synthetic time-series from $N=78$ neurons generated via homogeneous Poisson process which contains planted communities undergoing community events at every $\tau = 1000$ms. The data is divided into 1000ms windows and six functional network snapshots representing the co-activity of neurons are constructed by calculating the maximum cross-correlation between pairs of spike trains. 
		\textbf{B.} A two-snapshot dynamic network in which temporal carryovers are performed via either \emph{interlayer edges} by MMM and Infomap (left) or by some \emph{fixed rule} by DSBM, DPPM and Tensor Factorization (right). 
		\textbf{C.} Three different scenarios for the community events taking place in part B. On the left, two planted communities in $t$ `merge' so that the resulting community in $t+1$ has a new community label. In the middle, planted community I `grows' by joining with planted community II, and the resulting community has the same label as I. On the right, planted community II `grows' by joining with planted community I, and the resulting community gets the label II. Different DCD methods handle these types of carryovers differently.
	}
	\label{fig1}
\end{figure}

Neuronal systems are a prime example of a dynamic system that can be modeled as a temporal network.  For example, spike train data describes the simultaneous firing patterns of neurons over time.  Thus, one can build a functional network whose nodes are neurons and whose edges represent statistical relationships (such as synchronization or some other similarity measure) between the firing patterns of neurons.  In order to capture the fact that interactions between pairs of neurons will change over time, a common way of building a temporal network with this data is to create sequential snapshots of the network over time that describe the dynamic evolution of the data.  To do this, one can split the time series into smaller time series, construct chronologically ordered set of network states, and try to characterize the intrinsic patterns of connectivity across those individual snapshots (Fig.\ref{fig1}A).

One aspect of temporal networks that is often of interest to study is the dynamic properties of communities within the network over time (i.e., how communities might be born or die as a function of time and how nodes change community membership as the network structure evolves).  In our example of neuronal firing, communities could represent synchronized groups of neurons (cell assemblies), and we could ask how the membership of such groups changes over time.  Multiple dynamic community detection (DCD) methods have been developed that extend static community detection to temporal networks, where now communities can exist (and be created/die) across time \cite{dyncomdet1}.  However, similar to the case of static networks, each DCD method is based on a different definition of how communities are detected within the network.  Further, DCD methods must also include a definition of how to carry-over or assign community labels across snapshots (layers of the network).  As a result, DCD methods in the literature vary greatly depending on their treatment of the snapshots and their temporal dependence \cite{survey}. Some methods treat individual snapshots separately, others might iterate over the snapshots in chronological order, and some might use interlayer edges to link the snapshots over time into a temporally connected network.

Here, we focus on five commonly used DCD methods that span the different ways of defining dynamic communities:  Multilayer modularity maximization (MMM) \cite{MMM}, Infomap \cite{mapeq,infomap}, Dynamic stochastic block model (DSBM) \cite{dsbm,dsbm2}, Dynamic plex propagation method (DPPM) \cite{dppm}, and Tensor Factorization \cite{tensorfact}. MMM and DSBM define a community as a densely connected cluster of nodes with respect to a null model, whereas Infomap defines a community as a group of nodes in which information flows quickly and efficiently. DPPM utilizes a definition in which communities are groups of subsets (plexes) of fully connected subgraphs (cliques) that have maximal overlap. Finally, Tensor Factorization takes an approach from linear algebra and defines the communities as the bases of a vector space generating the underlying network. 

This variance in the definitions of a dynamic community forces these methods to make specific assumptions about how to temporally carry-over community labels across snapshots (layers) (Fig.\ref{fig1}B). Methods like MMM and Infomap operate on the idea that temporal carry-over is performed through the structural multilayer network topology; in this case `interlayer edges' are defined that link nodes across layers, such that communities can naturally exist across time. However, the other three methods use `fixed rules' to define temporal carry-over that ignore data-specific differences. DSBM uses a fixed generative model for the temporal network in which communities are created and transferred across time via a Bayesian algorithm. DPPM uses a fixed algorithm in which plexes in static layers are carried over across time if they intersect sufficiently between snapshots.  Finally, Tensor Factorization utilizes a fixed factorization algorithm (PARAFAC) that splits the 3-way tensor into simpler matrices in which the time component of the factorization corresponds to the temporal carry-over. 

Importantly, because of the different ways in which each method defines a community, both statically and dynamically, different methods will emphasize different features of the data and therefore will detect different patterns of dynamic communities.  It is therefore essential to have an understanding of how each method detects the specific features of the data and incorporates this information into the detected communities.  This is especially relevant in order to interpret any results when these methods are applied to experimental data sets where the underlying ground truth is not known.  Motivated by our example from neuroscience, here we are especially interested in how various DCD methods perform to detect data with a high presence of singleton communities (independently firing neurons) and transient communities (cell assemblies that change over time with the state of the brain).  We therefore simulate spike train data with known community structure and test the performance of DCD methods on this data.  

As expected, we find that different methods detect different patterns of dynamic community structure for the same data set.  Methods that incorporate interlayer edges to link snapshots over time perform better at detecting singleton and transient communities in our simulated data, but all methods struggle to perform temporal carry-over of community labels.  We find that the topology of how interlayer links are defined in these temporal networks can greatly influence the performance of the method.  The most common technique of interlayer coupling, called diagonal coupling, in which network nodes are linked to themselves in sequential layers, performs poorly at assigning the carry-over of community labels in our data.  However, we find that by utilizing information about the \emph{intralayer} topology of each individual layer in the network to couple the layers, one can improve DCD performance.  

Using techniques from topological data analysis (TDA), a field in the intersection  of data science and algebraic topology in mathematics \cite{tdaroadmap,TDA1,TDA2,TDA3}, we define a novel interlayer coupling method called skeleton coupling that defines interlayer edges based on the community information within the static layers of temporal networks.  Skeleton coupling takes the temporal neighborhood history and community assignment of a vertex (in the adjacent past state) into account such that performance of DCD methods are improved in terms of the temporal carry-over of both singleton communities and larger assemblies. We compare our results for skeleton coupling with previously proposed mechanisms of interlayer coupling and show that skeleton coupling outperforms other methods on data with a high prevalence of singleton and transient communities.

\section{Simulation of neuronal data}

As previously mentioned, this work is motivated by applications for studying temporal functional networks built from firing patterns of individual neurons.  In these functional networks, nodes are neurons, and the connections are given by calculating correlations between pairs of neuronal firing patterns.  Thus, in this type of network data, we expect to see many singleton communities (neurons whose firing patterns are not correlated with other neurons) as well as the presence of multiple transient communities (different synchronized cell assemblies that are born and die as neurons transiently fire together to perform computations).  However, because in real-world neuronal data the ground truth of community evolution cannot be known, here we apply our analysis to synthetic data.  Although previous work studying community evolution has designed benchmark networks for testing community detection in evolving networks \cite{benchmarks1,benchmarks2,benchmarks3,benchmarks4}, the links in these networks represent structural connections between nodes.  As such, these benchmark models do not generally contain singleton communities or highly transient communities as commonly seen in functional networks based on correlation data \cite{comcorr1,comcorr2,comcorr3} like in the case of functional neuronal networks.  We therefore designed a set of functional benchmark networks built from correlations between simulated neuronal spike trains similar to the procedure described in \cite{localupd}.  See the Methods section for details of how time series of spike trains were simulated and correlated.

In our numerical experiments, we study two different types of community events expected to be present in dynamic functional networks: monotonic and non-monotonic events.  Monotonic events correspond to the scenarios where a graph progressively evolves over time such that communities in one layer are nested into the communities in an adjacent layer.  (Recall that in dynamic community assignment, communities are composed of node-layer elements because nodes can change communities from layer to layer.) Mathematically, we call a community evolution scenario monotonic if community $p$ at time step $t$, $C_{p}^{t}$, satisfies $C_{p}^{t}\subseteq C_{p}^{t+1}$ (or $C_{p}^{t+1}\subseteq C_{p}^{t}$ for time reversed) for all communities and time steps in the temporal network.   Such events include community expansion, shrinkage, or continuation.  Non-monotonic evens represent the scenarios in which communities in adjacent layers can partially overlap (as in Fig.\ref{fig1}A), but these communities do not necessarily contain each other. Examples of non-monotonic events include community merging, splitting, death, and birth.  In each of these scenarios, it is necessary to determine how the community labels should evolve over time, as depending on the properties of the data (such as neuronal firing pattern or rate), one might want to either define a new community, or carryover a previous label (see Fig.~\ref{fig1}(B-C)).

Here, we focus on examples of a monotonic event (an expanding community) and a non-monotonic event (multiple transient communities). Community structure is modeled using simulated neuronal spiking activity with built-in correlations between firing patterns of individual neurons within a given community.  In addition, the community structure (correlated firing of neurons) is allowed to dynamically evolve through a series of community events. Importantly, in this data, multiple neurons have independent firing patterns, such that many singleton communities are present in the data.

In order to map this data to a temporal network, the time series is first divided into multiple windows, each representing a layer of the network.  Functional network structure in each layer is obtained by computing the absolute value of the pairwise maximum cross-correlation between firing patterns of neurons over the window.  Because the use of cross-correlations to define functional network connections results in a fully connected network with many small edge weight values that likely represent noise in the data, for each data set, we create a set of temporal networks in which a threshold is used to eliminate connections with edge weights below the threshold value.  In the following section, all results are presented across a range of threshold values (shown along the x-axis in the parameter space maps of Figs. \ref{fig2}, \ref{fig5}, and \ref{fig6}).  Please see Methods for further details of synthetic data generation and network creation.

\section{Comparison of DCD method performance}\label{section_dcd}

We compare the performance of 5 different DCD methods (MMM \cite{MMM}, Infomap \cite{multimapeq}, DSBM \cite{dsbm}, DPPM \cite{dppm} and Tensor Factorization \cite{tensorfact}) on two different community evolution scenarios as described above (expanding and transient communities). We include a range of method specific hyperparameters (resolution parameter $\gamma$, multilayer relax rate $\rho$, degree correction $\Delta$, k-plex dimension $k$, and input tensor rank $\eta$, respectively) which are varied across the y-axes of parameter space maps for these 5 methods. We note that while these hyperparameters are not comparable across methods, the user must make a choice of each parameter when implementing the algorithm, thus we vary the parameter to show the influence of user choice.

\begin{figure}[!h]
	\centering
	\includegraphics[width = 1\linewidth]{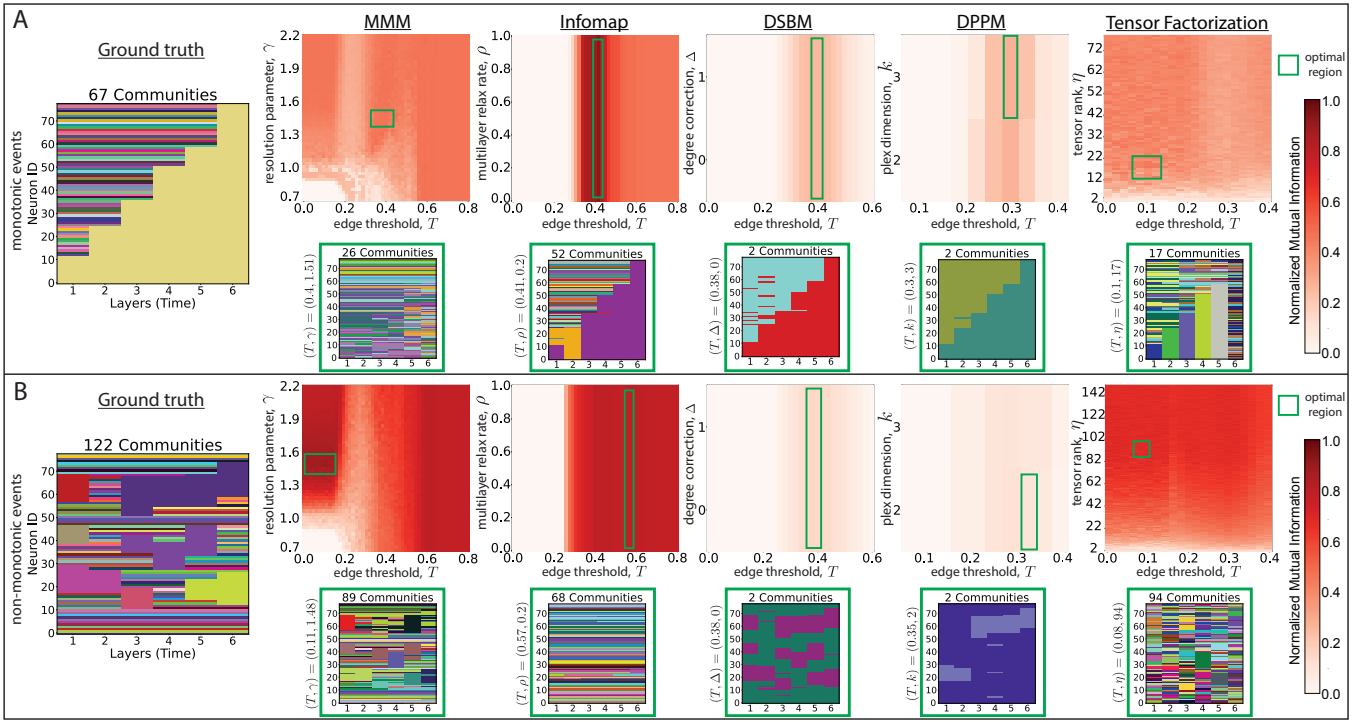}
	\caption{
		\textbf{Comparing DCD methods on simulated time series.} 
		A comparison of method performance across parameter spaces and \emph{example partitions} for the optimal regions of five different DCD methods: MMM, Infomap, DSBM, DPPM and Tensor Factorization. Method performance is explored as a function of the edge threshold, $T$, and a method-specific parameter (resolution parameter $\gamma$, multilayer relax rate $\rho$, degree correction $\Delta$, k-plex size $k$ and tensor rank $\eta$, respectively) by plotting the normalized mutual information (NMI) between the `ground truth' community labels (left panel) and predicted labels. The optimal region is defined as the parameter values $(T,\cdot)$ in which NMI is maximized, and an \emph{example partition} from this region is shown under each parameter grid. Partition plots represent the community evolution across six snapshots (network layers) and the colors indicate the community label of each neuron at each point in time. 
		\textbf{A.} Dynamics of $N=78$ spiking neurons are simulated such that a large community keeps expanding by merging with singleton communities at every layer (monotonic event). There are 67 community labels in total during the ground truth community evolution.
		\textbf{B.} Dynamics of $N=78$ spiking neurons are simulated such that synchronized groups of neurons appear and disappear over time i.e. \emph{transient communities}, and neurons that are not part of any community are assigned a unique community label (indicated by colors) that is temporally carried over unless a neuron undergoes a community event. A total of 122 community labels are produced during this event as shown in the ground truth partition plot.
	}
	\label{fig2}
\end{figure}

In the left panels of Fig.\ref{fig2}A and Fig.\ref{fig2}B, we display the ground truth of the community evolution of planted dynamic communities.  We then plot the parameter space describing the performance of the method as a function of the normalized mutual information (NMI) \cite{nmi1,nmi2} with respect to the ground truth (See Methods Section `Evaluating partition quality').  In these plots, the parameter values representing the optimal performance of each method are indicated by the region bounded by the green rectangle (See Methods Section `Optimal regions').  An example of the community evolution in this optimal regime is shown below the parameter space plot.

In Fig.~\ref{fig2}(A), we present the performance of the 5 DCD methods to detect the community evolution of an expanding community event.  Neurons first exist as singleton communities (firing patterns are uncorrelated with others) and join a growing correlated community as time advances (series of monotonic events). As seen in the NMI parameter landscapes, each method varies in its ability to better detect this pattern of community evolution.  Example community evolution plots and respective method hyperparameters are shown for the optimal method performance below these plots.  It can be observed that the MMM and Infomap methods perform the best at detecting singleton communities and performing temporal carryover; these methods also produce the highest NMI values (darker shade of red) over a wider range of method hyperparameters.  Still, MMM fails to detect the expanding community, whereas Infomap partially detects this growing community, albeit with some noise.  DSBM and DPPM, on the other hand, yield relatively low NMI and result in the detection of 2 total communities, as they do not distinguish the singleton communities and instead lump all uncorrelated neurons into a single community.  Tensor Factorization performs somewhere in-between these extremes and detects most of the communities in individual layers separately, failing to perform temporal carryover.

We next compare the performance of the methods on data containing transient and singleton communities (non-monotonic events; Fig. \ref{fig2}B).  Again, we observe that MMM and Infomap perform the best as measured by the NMI, but the optimal community partitions shows that they are detecting rather different patterns of community evolution.  Both methods are able to detect singleton communities and perform temporal carryover on the singleton communities.  MMM additionally, detects some of the transient larger communities, but fails to perform temporal carryover between layers for these transient communities.  Tensor Factorization can also detect the transient communities in addition to singleton communities but completely fails at performing temporal carryover of community layers.  Once again, DSBM and DPPM only detect two communities which does not reflect the planted structure and is apparent in their low NMI values.

It is notable that in each of the scenarios studied, the DSBM and DPPM methods were unable to detect the presence of singleton communities in the data. It is worth observing that dynamic stochastic block models cannot capture singleton communities because, simply by definition, a community is in a block structure with multiple nodes for these methods, and singleton communities can not form blocks. Similarly, DPPM defines a community as a $k$-plex which is an assemble of multiple nodes. Therefore, singleton communities can not form plexes. Further, Tensor Factorization consistently failed to perform temporal carryover of detected communities, which is most likely an approximation artifact due to the solver's tolerance used in this method.  While MMM and Infomap did not always perform the temporal carryover, they were able to identify singleton communities, and importantly, these methods rely on interlayer edges to link layers across time.  In the data presented above, for the MMM and Infomap methods, the standard technique of diagonal coupling was used, as this is the most commonly employed technique of coupling.  However, this is another parameter that can be tweaked when using these methods, and for the remainder of this paper, we will focus on the use of different techniques of interlayer coupling to further improve community detection in the MMM and Infomap methods.

\section{Traditional interlayer coupling}

\begin{figure}[!h]
	\centering
	\includegraphics[width = \linewidth]{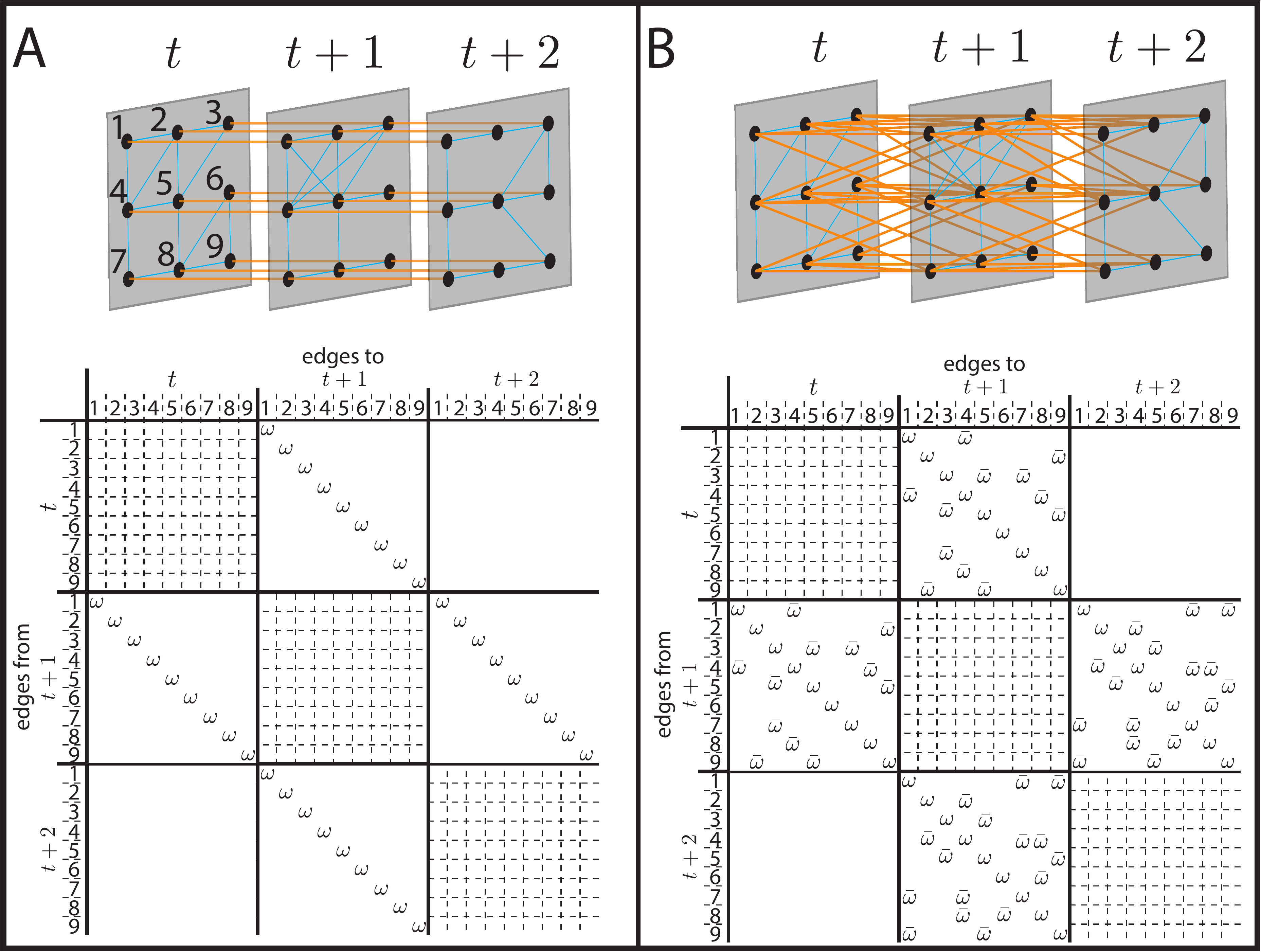}
	\caption{
		\textbf{Types of interlayer coupling heuristics.} 
		\textbf{A.}  Diagonal coupling. Nodes are only allowed to be connected to their past and future self with a uniform edge weight $\omega$ (uniform coupling). A modification to this (local updates) can be done by dynamically altering the edge weights based on a self-similarity metric of the node from time $t$ to $t+1$. 
		\textbf{B.} Non-diagonal coupling. Interlayer coupling can exist between any pair of nodes. In the example, we illustrate neighborhood coupling which connects neighborhoods of every node with the adjacent layer with edge weights of $\omega$ and $\bar{\omega}$. See the Methods section `Interlayer coupling' for details.
	}
	\label{fig3}
\end{figure}

When employing the MMM and Infomap methods in a temporal network, the network must be constructed by using interlayer edges that describe how nodes are linked across layers (and therefore across time).  Here, we review some standard and recently proposed techniques for defining interlayer links in temporal networks.  Let $\mathsf{T} = (V,E)$ be a node-aligned temporal network with snapshot representation $\mathsf{T} = \{\mathsf{G}^{1},\mathsf{G}^{2},...\mathsf{G}^{t_{max}}\}$. We'll denote the node $i$ in the layer $\mathsf{G}^{t}$ by $i^{t}$ and an edge from $i^{t}$ to another node $j^{s}$ by $(i^{t},j^{s})$. Note that we will only be focusing on ordinal coupling (i.e., nodes are allowed to be connected only to other nodes in adjacent layers) throughout the rest of the paper.

\subsection{Diagonal coupling}
The most common way of constructing interlayer edges is to use diagonal coupling.  In this case, each node of the network is coupled with its temporal counterpart in a regular fashion as in Fig.\ref{fig3}A. Thus, there exists an interlayer edge $(i^{t},i^{t+1})$ for all $i\in V$ and for all $t\in \{1,2,..,t_{max}-1\}$ with edge weight $\omega$, where $\omega$ is constant across all edges, but its value must be specified by the user.  Here, $\omega$ can be thought of as a self-identity link that preserves the identity of the node throughout time.  We will refer to this technique of coupling as \textit{uniform diagonal coupling}.

While this technique of coupling will allow for a node to maintain its identity throughout the network, it does not capture the fact that in many networks, the nodes represent dynamic entities whose properties change throughout time.  A question one may ask is if changing the values of the interlayer edge weights (i.e., allowing for $\omega$ to vary across nodes and layers) would make a difference in the detection of dynamic communities.  Each diagonal interlayer edge is a link from a node to its future or past self, so in this sense, these links indicate the strength of temporal self-similarity of nodes.  In our simulated data, the nodes are in fact neurons whose firing rates and patterns can evolve, thus a nodes self-similarity  over time is not necessarily constant.  Previous work \cite{localupd} has described a technique that allows for the value of $\omega$ to change based on the level of nodal self-similarity.  The greater the change in the self-similarity of a node is between snapshots (e.g. the firing rate of the neuron), the weaker the nodes interlayer edge weight is between the corresponding temporal layers. We refer to this technique as \textit{diagonal coupling with local updates} and, mathematically, we assign an interlayer edge between $i^{t}$ and $i^{t+1}$ for all $i \in V$ and for all $t\in \{1,2,..,t_{max}-1\}$ with edge weight $\omega_{i}^{t}$ depending on the spike rate change in node $i$ from $\mathsf{G}^{t}$ to $\mathsf{G}^{t+1}$.  See the Methods Section `Interlayer coupling' for details.

\subsection{Non-diagonal coupling}

While diagonal coupling only allows for a link between a node and itself across layers of the network, it is also completely reasonable to relax this restriction and allow links between some or all pairs of nodes, which introduces a new dimension of complexity and increases the size of the parameter space enormously.  While there are multiple ways that one could perform a non-diagonal coupling scheme, here we highlight one previously proposed algorithm called \textit{neighborhood coupling} \cite{neigborhoodcpl}, that connects a maximal neighborhood around every node with the adjacent layers (Fig.\ref{fig3}B).  Mathematically, we assign interlayer edges of constant weight $\omega$ from $i^{t}$ to a set $\{j^{t+1}\}_{j \in \mathsf{N}_{i}^{t}}$ such that $j$ is in the maximal neighborhood of $i^{t}$ in terms of edge weight (strongly connected neighbors of $i$ in $\mathsf{G}^{t}$) where $\mathsf{N}_{i}^{t} = \{j^{t}|(j^{t},i^{t})\in E_{t}\}$. See the Methods Section `Interlayer coupling' for details.  This algorithm of coupling is based on the assumption that the topology of the network in its previous state affects how the network evolves. Note that this algorithm also results in a more dense coupling than that of diagonal coupling as seen in Fig. \ref{fig3}B. Importantly, while this method creates interlayer edges based on the intralayer topology of layer $t$ around node $i$, such topology does not necessarily encode the community assignments of (i,t) in the interlayer topology, as nodes can be connected to other nodes in different communities.  Regardless, prior work has indicated that the use of neighborhood coupling can aide in the detection of transient communities \cite{neigborhoodcpl}.

\begin{figure}[!h]
	\centering
	\includegraphics[width = 1\linewidth]{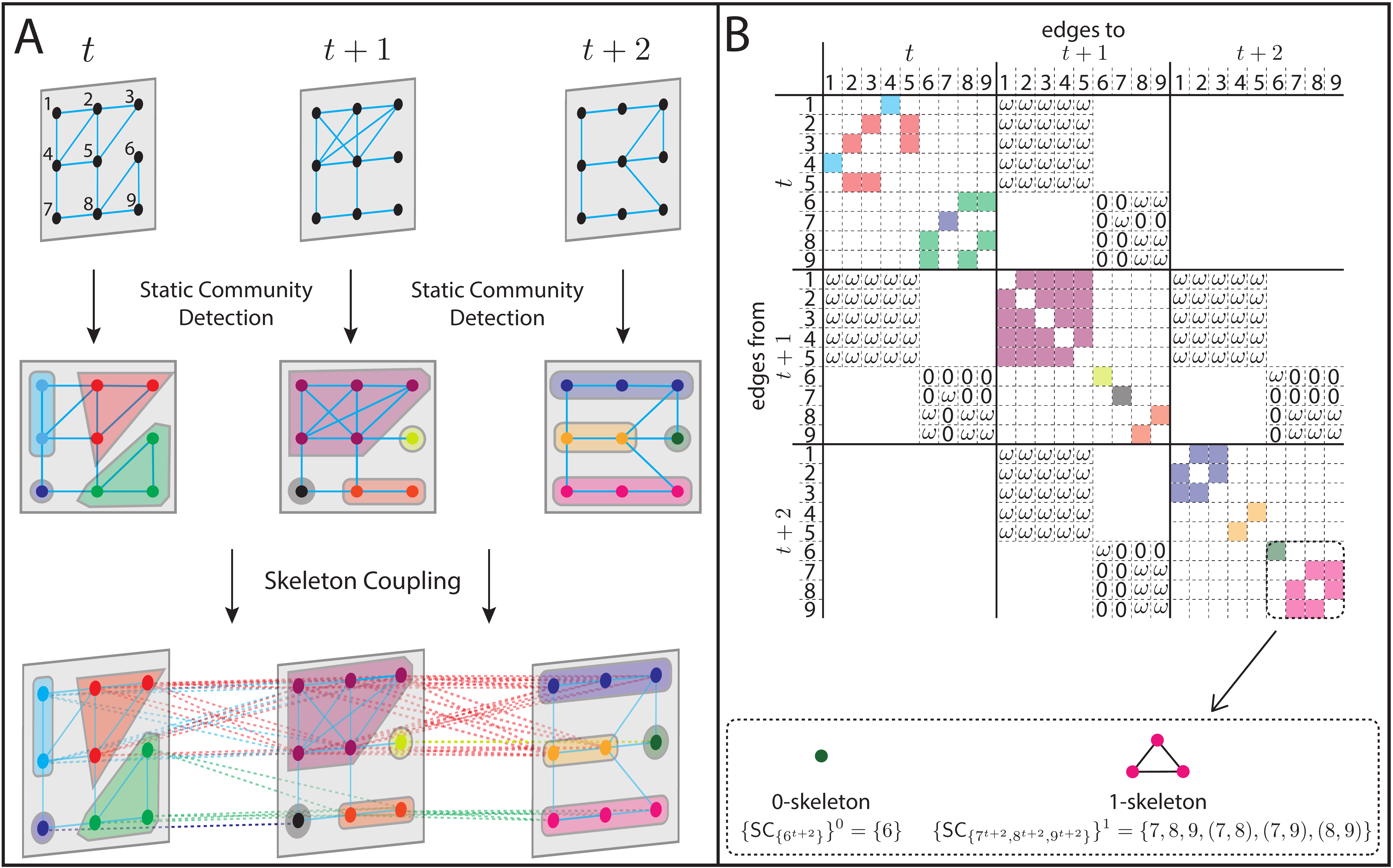}
	\caption{
		\textbf{Skeleton coupling.}
		\textbf{A.} Schematic of our proposed skeleton coupling framework for determining non-empirical interlayer edges in dynamic community detection. Static community detection is performed on individual layers to find the communities in each layer (indicated by colors). Interlayer edges are then assigned via skeleton coupling, and dynamic community detection is applied to the resulting temporal network.
		\textbf{B.} Supra-adjacency matrix of the toy example in \textbf{A.}. Colors represent the $0$- or $1$-skeletons formed by the nodes in the corresponding communities found by the primary static community detection. For example, the color of the entry $\mathsf{G}^{t+2}_{(6,6)}$ represent the $0$-skeleton on the vertex set $\{6^{t+2}\}$. Alternatively, the colors of the entries $\mathsf{G}^{t+2}_{(7,8)}, \mathsf{G}^{t+2}_{(8,7)}, \mathsf{G}^{t+2}_{(7,9)}, \mathsf{G}^{t+2}_{(9,7)}, \mathsf{G}^{t+2}_{(8,9)}$ and $\mathsf{G}^{t+2}_{(9,8)}$ represent the $1$-skeleton on the vertex set $\{7^{t+2},8^{t+2},9^{t+2}\}$.  We also display these examples of skeletons drawn as a simplicial complex with corresponding mathematical descriptions at the bottom of the panel.  Note that, skeleton coupling is a type of uniform, ordinal and non-diagonal coupling. In order to determine the interlayer edges between layers $\mathsf{G}^{t}$ and $\mathsf{G}^{t+1}$, we compare the skeletons that vertices constitute. See the text for descriptions of interlayer edges between layers.
	}
	\label{fig4}
\end{figure}

\section{Skeleton coupling}

While neighborhood coupling has the advantage of incorporating the topology of the current state of the network into the coupling between layers, this approach does not directly address our desire to improve community carryover between layers, as the neighborhood of a node is distinct from its community assignment within that layer.  We therefore propose a novel algorithm of non-diagonal coupling that we call \textit{skeleton coupling} that is designed to link network layers based on the static community structure within layers, therefore promoting the correct temporal carryover of community labels.

The main idea of skeleton coupling is to assign interlayer links to a node either sparsely or densely depending on its temporal neighborhood history creating temporal channels between snapshots. Moreover, by definition of a dynamic community, interlayer coupling links should only exist between the temporally carried over communities because these links are directional maps (due to the asymmetric nature of time), mapping a previous state of the system into a future state. We therefore start by finding the static communities in every snapshot (network layer) in order to determine the domain and range of these maps. That is, we find the partitions $\mathsf{P}_{t} = \{C_{1}^{t},C_{2}^{t},C_{3}^{t},...\}$ for all snapshots $\mathsf{G}^{t}$ of a temporal network $\mathsf{T}$, where $C_{p}^{t} = \{i^{t}|\exists i \in V \}$ consists of nodes of the network belonging to the community $p$ obtained from a static community detection method at that layer $t$. This step could be performed using any static community detection method, although here we use the same static version of the DCD method applied later (Fig.\ref{fig4}A).

Next, we will define a $k$-skeleton of a community. We first define a $k$-simplex, $(0,1,...,k)$, as an ordered set of vertices with cardinality $k+1$, e.g., a $0$-simplex is a vertex $(i)$ (or simply $i$) and a $1$-simplex is an undirected and an unweighted edge $(i,j)$. We define a $K$-dimensional simplicial complex of a community $C_{p}^{t}$ as the union of all possible simplicies on the vertex set $C_{p}^{t}$ up to dimension $K$, that is $\{\mathsf{SC}_{C_{p}^{t}}\}_{k=0}^{K}$. Lastly, a $k$-dimensional skeleton of a simplicial complex of a community (or simply a $k$-skeleton of a community) is the union of all simplicies on the vertex set $C_{p}^{t}$ up to dimension less than or equal to $k$, i.e., $\{\mathsf{SC}_{C_{p}^{t}}\}^{k}$. See the bottom of Fig.\ref{fig4}B) for some examples.

By focusing on the case $k\leq 1$ for computational simplicity, it's important to note that a community needs $k+1$ elements to be able to form a $k$-skeleton. For example when $k=1$, $|C_{p}^{t}|$ needs to be greater than or equal to $2$. Therefore, singleton communities, i.e., $|C_{p}^{t}|=1$ cannot form $1$-skeletons. The crux of skeleton coupling algorithm relies on the fact that $1$-skeletons and $0$-skeletons are  topologically different objects (since one has edges in it and one don't), we do or do not assign interlayer links between these two objects across layers accordingly.

In particular, for any node $i^{t}$ in layer $\mathsf{G}^{t}$, there are then two possibilities. The first one is $i^{t}$ can belong to a community of size greater than or equal to 2, i.e., $|C_{p}^{t}|\geq 2$ such that $i^{t}\in C_{p}^{t}$. In this case, $i^{t}$ is part of a 1-skeleton $\{\mathsf{SC}_{C_{p}^{t}}\}^{1}$). The second one is that $i^{t}$ can be a singleton community, i.e., $|C_{p}^{t}|=1$ such that $\{i^{t}\} = C_{p}^{t}$. In this case, $i^{t}$ has a 0-skeleton $\{\mathsf{SC}_{\{i^{t}\}}\}^{0}$). Next, we look at $i^{t}$'s counterpart in the layer $\mathsf{G}^{t+1}$ to see if $i^{t+1}$ is part of a $0$ or $1$-skeleton.  In order to determine how to design the interlayer coupling, we then consider all $4$ possibilities of combinations between the skeletons of $i^{t}$ and $i^{t+1}$:

\begin{description}
	
	\item \textbf{case i:} \textit{$i^{t}$ and $i^{t+1}$ are both $0$-skeletons:}
	We assign an undirected interlayer edge $(i^{t},i^{t+1})$ with uniform edge weight $\omega$ (i.e. we diagonally couple $0$-skeletons). This situation, in general, describes the continuation of community label that is maintained by the same singleton over time. This is seen in Fig.~\ref{fig4}B, where $7^{t}$ and $7^{t+1}$ are both $0$-skeletons ($\{\mathsf{SC}_{\{7^{t}\}}\}^{0}$ and $\{\mathsf{SC}_{\{7^{t+1}\}}\}^{0}$ indicated by dark blue and black squares, respectively)–  there is a bidirectional self-identity link $(7^{t}, 7^{t+1})$ with weight $\omega$ between them. Similarly, $6^{t+1}$ and $6^{t+2}$ are also linked by a single self-identity edge.
	
	\item \textbf{case ii:} \textit{$i^{t}$ is a $0$-skeleton and $i^{t+1}$ is part of a $1$-skeleton:}
	In this case, we do not assign any interlayer edges from $i^{t}$ to the next snapshot $\mathsf{G}^{t+1}$ since we don't want a singleton community label to persist when the singleton node joins a larger community. For example, in Fig.\ref{fig4}B, observe that $7^{t+1}$ is a $0$-skeleton ($\{\mathsf{SC}_{\{7^{t+1}\}}\}^{0}$ indicated by black square), but $7^{t+2}$ is part of a $1$-skeleton ($\{\mathsf{SC}_{C_{p}^{t+2}}\}^{1}$ such that $C_{p}^{t+2} = \{7,8,9\}$ indicated by the cluster of pink squares). We therefore did not assign any interlayer links associated with this node between the two snapshots, as indicated by the weights $0$ in the corresponding entries of the supra-adjacency matrix.
	
	\item \textbf{case iii:} \textit{$i^{t}$ is part of a $1$-skeleton and $i^{t+1}$ is a $0$-skeleton:}
	This case is the time-reversed version of \textit{case ii}. We do not assign any interlayer edges from the node $i^{t}$  to any node in time step $\mathsf{G}^{t+1}$ since this case describes a community shrinking and splitting, and we don't want the community label of the node $i^{t}$ to be carried over to the next time step. Note in Fig.\ref{fig4}B, for example, that $6^{t}$ is part of a $1$-skeleton ($\{\mathsf{SC}_{C_{p}^{t}}\}^{1}$ such that $C_{p}^{t} = \{6,8,9\}$ indicated by the cluster of green squares) and $6^{t+1}$ is a $0$-skeleton ($\{\mathsf{SC}_{\{6^{t+1}\}}\}^{0}$ indicated by the yellow square). We therefore did not assign any interlayer links associated with this node between the two snapshots, as indicated by the weights $0$ in the corresponding entries of the supra-adjacency matrix.
	
	\item \textbf{case iv:} \textit{$i^{t}$  and $i^{t+1}$ are both parts of $1$-skeletons:}
	We assign interlayer edges of uniform strength $\omega$ from every node with which $i^{t}$ shares a community $C_{p}^{t}$ to every other node with which $i^{t+1}$ shares a community $C_{q}^{t+1}$ in the snapshot $\mathsf{G}^{t+1}$. Depending on the sizes of the communities that $i^{t}$ and $i^{t+1}$ are part of, this situation can describe multiple types of community events. Regardless, we want the community label to persist over time. In Fig.\ref{fig4}B, notice for example, $3^{t}$ and $3^{t+1}$ belong to communities of size larger than 1 $C_{p}^{t}=\{2^{t},3^{t},5^{t}\}$ and $C_{q}^{t+1} = \{1^{t+1},2^{t+1},3^{t+1},4^{t+1},5^{t+1}\}$ (and the corresponding simplicial complexes $\{\mathsf{SC}_{C_{p}^{t}}\}^{1}$ and $\{\mathsf{SC}_{C_{q}^{t+1}}\}^{1}$ indicated by the clusters of red and purple squares, respectively). This implies we add bidirectional interlayer edges from $3^{t}$ to $C_{q}^{t+1}$ with weights $\omega$. If we look at other elements of $C_{p}^{t}$, $2^{t}$\&$5^{t}$ and their counterparts in the next layer $2^{t+1}$\&$5^{t+1}$, we see a similar scenario, and therefore, we add bidirectional links from all the nodes of the community $C_{p}^{t}$ to all of the nodes of $C_{q}^{t+1}$ with weights $\omega$, building a temporal bridge between them.
	
\end{description}

Skeleton coupling thus serves as a finely tuned coupling algorithm based on the topological difference between singletons and larger communities in the time-varying network.  Here, we use the terminology of a `$k$-skeleton' from topological data analysis (TDA) \cite{tdaroadmap, TDA1, TDA2, TDA3} because we claim that the definition of a dynamic community in functional networks is in the form of $k$-plexes \cite{dppm}.  In a perfect world of noiseless data, a community of $n$ nodes should be an $n$-clique, whereas in reality, a dynamic community is a set of nodes that has missing or noisy links.  We therefore rely on the simplicial complex definition and usage of skeletons to account for real-world data in which true cliques are unlikely to be present within communities. We also provide a pseudo-code of the implementation of skeleton coupling in blue Appendix A `Skeleton coupling algorithm'.

\section{Applications of skeleton coupling to temporal network analysis}

In Section \ref{section_dcd}, we showed that the MMM and Infomap methods were capable of detecting singleton communities within the data, but performed poorly in carrying over the correct community labels.  In these previous comparisons, we focused on the effect of the resolution parameter, $\gamma$, (MMM) and the multilayer relax rate, $\rho$, (Infomap) as a function of the edge threshold value, $T$.  Here, we will now fix $\gamma$ and $\rho$ and instead explore the effects of incorporating four different interlayer coupling strategies: uniform diagonal coupling, diagonal coupling with local updates, neighborhood coupling, and our newly proposed skeleton coupling.  In each case, we explore the performance of the DCD method as a function of the interlayer edge weight, $(T,\omega)$, and edge threshold value value, $T$, which we measure by averaging the normalized mutual information ($<NMI>$) between the true and predicted labels across multiple stochastic runs of the algorithm. Below the NMI plots, we show the standard deviations of NMI, $\sigma(NMI)$, for each parameter pair $(T,\omega)$ across these stochastic runs. Also, see Appendix D `Maximum overlapping distances between stochastic runs' for additional results. As before, we compare method performance for two distinct types of community evolution scenarios: a monotonic series of community events in which an initial community expands over time until the whole network is synchronized, and a non-monotonic event in which \emph{transient communities} appear/disappear over time. In Appendix B `Further community evolution scenarios', we present very similar results for two additional set of community evolution scenarios in which we keep the number of neurons and layer the same, and vary size of the communities in each layer.   Note that we did not perform any experiments in which the number of layers was varied, as other previous work has explored how the number of layers can affect DCD performance \cite{muchaetal,multimapeq}.

We now show how incorporating the use of skeleton coupling into the design of networks results in improved performance of temporal carryover for both MMM and Infomap methods.

\subsection{Influence of skeleton coupling with MMM}

\begin{figure}[!h]
	\centering
	\includegraphics[width = 1\linewidth]{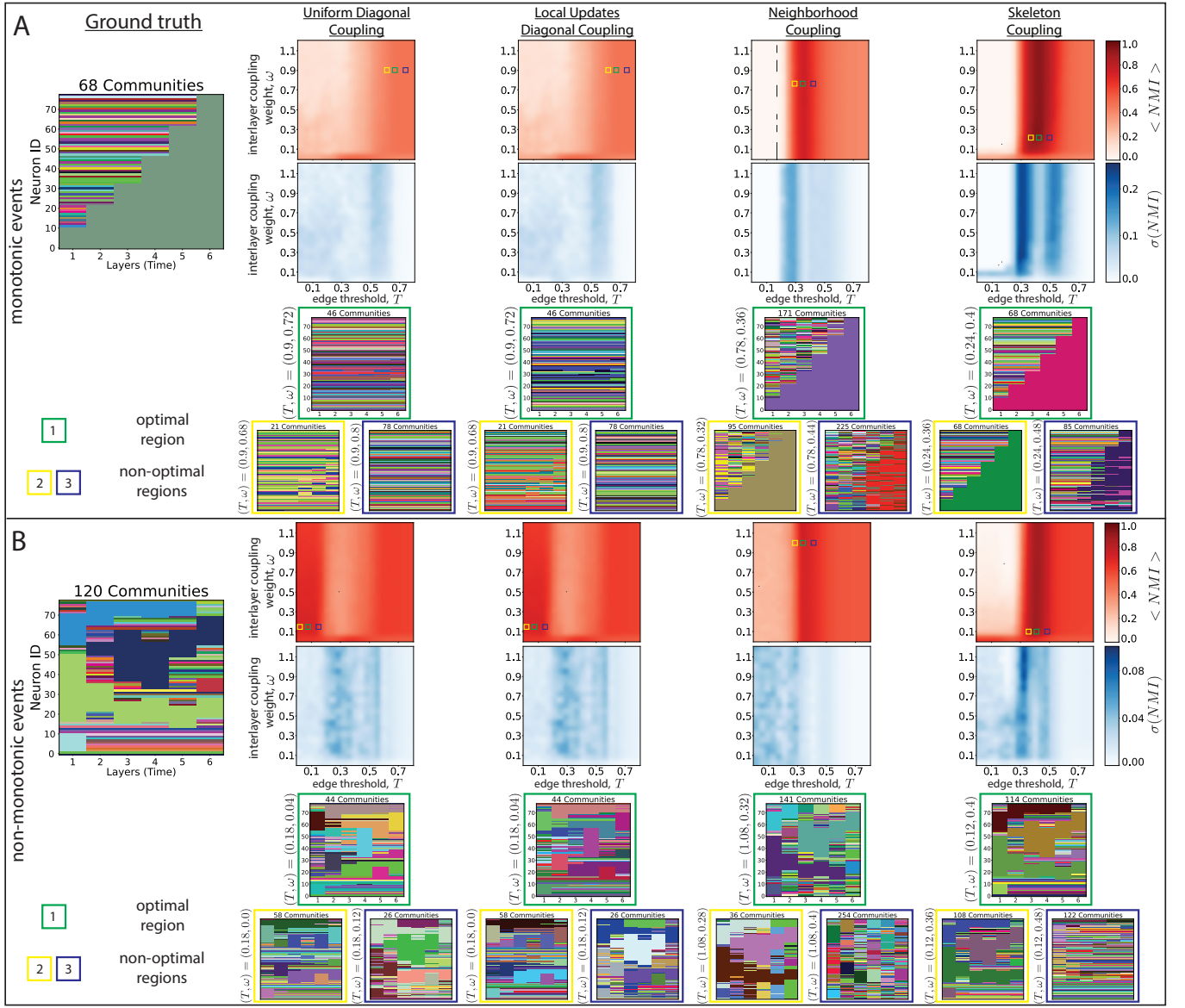}
	\caption{
		\textbf{Comparison of different interlayer coupling strategies for MMM.}
		Comparison of different interlayer coupling strategies for $N=78$ neurons undergoing \textbf{A.} expanding community events with 68 total community labels and \textbf{B.} transient community events with a total of 120 community labels. Community labels are depicted by color. Parameter landscapes show method performance measured by the average NMI and standard deviations across multiple stochastic runs in the $(T,\omega)$ parameter space.  The configuration model was chosen as the null model and the resolution parameter was equal to $0.98$ for the expanding events in A and $1.3$ for transient events in B. Under each parameter space, we illustrate \emph{example consensus partitions} found by the MMM method within the bounds of optimal and non-optimal regions highlighted by green, yellow and blue rectangles, in the order of descending average NMI, respectively. 
	}
	\label{fig5}
\end{figure}

We first compare the performance of MMM under different interlayer coupling strategies. In Fig.~\ref{fig5}(A), we present the results of using uniform diagonal coupling and diagonal coupling with local updates for an expanding community. Below the $<NMI>$ and $\sigma(NMI)$ plots, we show the community evolution plots obtained by the consensus partition \cite{consensusdissesnsus} across all stochastic runs of both optimal and non-optimal parameter choices.  Observe that both coupling techniques yield structurally similar results, failing to identify the expanding community and instead carry over the independent community labels. This result is true for both optimal and non-optimal choices of coupling parameters.

Next, we study non-diagonal coupling approaches using neighborhood coupling and skeleton coupling.  Notice that both of these non-diagonal approaches succeed in detecting the expanding planted dynamic community. However, neighborhood coupling fails to temporally carry over the singleton communities, assigning a different community label to each node at every layer, resulting in a total of 171 communities. On the other hand, skeleton coupling carries over the singleton communities better than other methods and results in a total number of detected communities that is within reasonable range of the ground truth. We also see an overall improvement in the performance of the method for non-optimal parameter regions: parameter regions that are near optimal (yellow boxes) can partially recover the expanding community.

In Fig.~\ref{fig5}(B), we next examine the performance of different coupling strategies in detecting transient communities in the data.  Notice that both diagonal coupling techniques perform relatively well in finding planted \emph{transient communities} within a layer. However, while some singleton community labels are temporally carried between layers, the diagonal coupling techniques perform poorly in carrying over the transient community labels between snapshots.

When using non-diagonal coupling algorithms, we observe that by both neighborhood coupling and skeleton coupling perform well in detecting and carrying over transient community labels.  However, neighborhood coupling performs poorly at carrying over independent community labels which results in a high number of total detected communities.  Skeleton coupling results in not only better detecting and carrying over transient communities, but is also able to improve carrying over independent community labels, resulting in higher performance values. Overall, skeleton coupling outperforms the other coupling heuristics when combined with MMM for both community evolution scenarios. Moreover, we claim that these results are robust for optimal partitions across different stochastic network generations as shown by the standard deviation plots under each parameter space. One can observe that for the optimal parameter choices $(T\omega)$, standard deviations of NMI is low between different stochastic runs. Lastly, all of these results hold for similar community evolution scenarios which we present in Fig. \ref{supfig0}.

\subsection{Influence of skeleton coupling with Infomap}

\begin{figure}[!h]
	\centering
	\includegraphics[width = 1\linewidth]{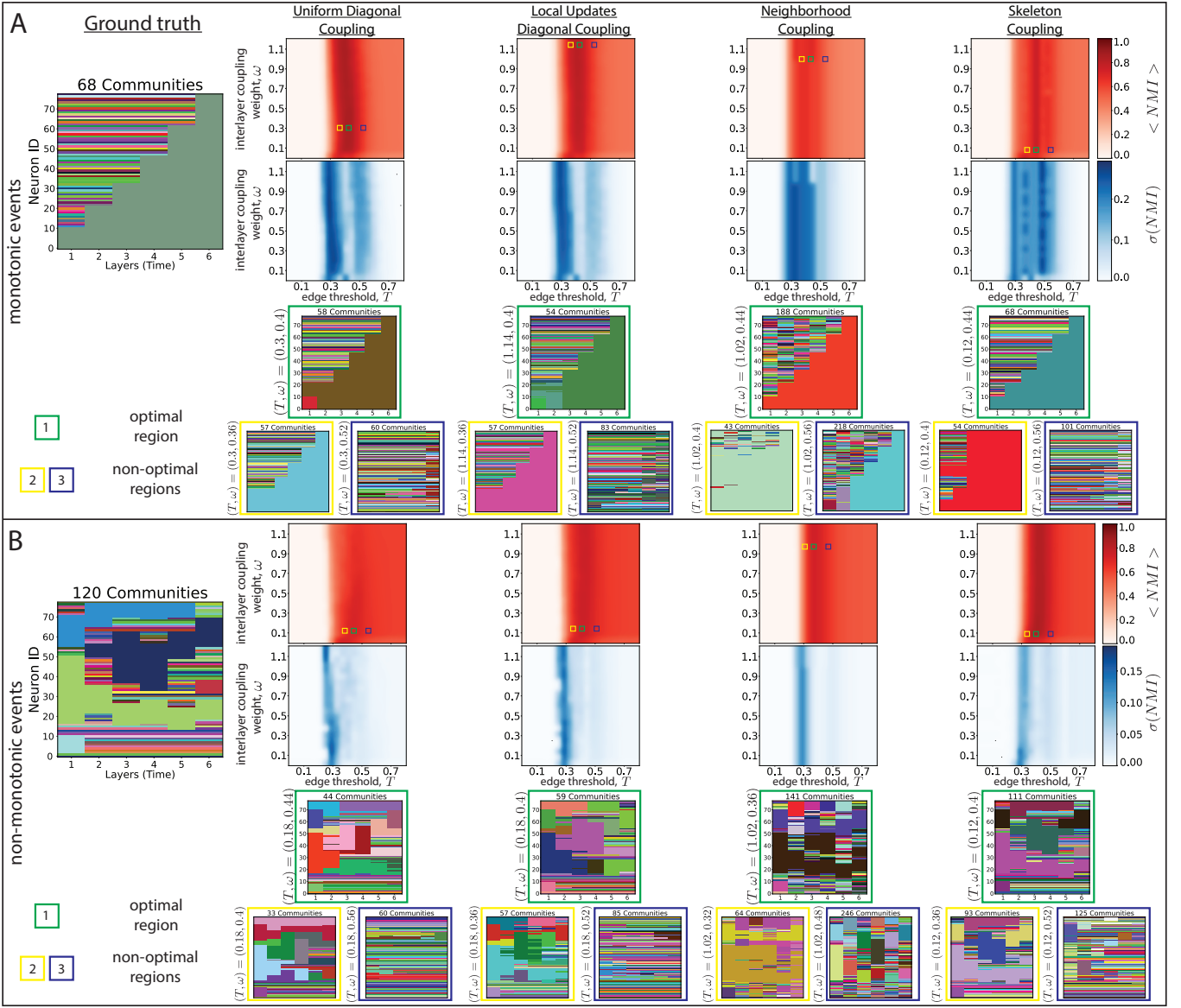}
	\caption{
		\textbf{Comparisons of different interlayer coupling strategies for Infomap.}
		Comparison of different interlayer coupling heuristics for $N=78$ neurons undergoing \textbf{A.} expanding community events with 68 total community labels and \textbf{B.} transient community events with a total of 120 community labels. Community labels are depicted by color. Parameter landscapes show method performance measured by the average NMI and the standard deviations in the $(T,\omega)$ parameter space. The multilayer relax rate was set to $0.2$ in both panels. Under each parameter space, we illustrate \emph{example consensus partitions} found by the Infomap method within the bounds of optimal and non-optimal regions highlighted by green, yellow and blue rectangles, in the order of descending average NMI, respectively. 
	}
	\label{fig6}
\end{figure}

In Fig.~\ref{fig6}, we show the performance of the four different coupling strategies when combined with the Infomap method for the same data as in Fig.~\ref{fig5}. Observe that in  Fig.~\ref{fig6}(A), for the expanding community, both diagonal coupling techniques seem to fail detecting the neurons contained in the first a few snapshots of the temporal network as part of the expanding community, possibly due to the number of snapshots in the temporal network. However, these coupling schemes perform fairly well at capturing the temporal carry over of singleton community labels.

For the case of the non-diagonal coupling schemes, both non-diagonal approaches identify the expanding community with high NMI. However, neighborhood coupling fails to temporally carry over the singleton communities, assigning each node a different community label in every layer. In contrast, skeleton coupling detects the temporal carryover of singleton communities better, resulting in a relatively good match between the detected evolution of communities and the ground truth.

Note that $<NMI>$ is higher (darker shade of red) for diagonal coupling approaches than non-diagonal coupling approaches. This seems to be an artifact of averaging the parameter spaces. Diagonal coupling approaches combined with Infomap happen to yield more consistent higher NMI across different stochastic runs for the same parameter values $(T,\omega)$. As a result, the average NMI is higher. On the other hand, this is not the case for non-diagonal coupling approaches, particularly for skeleton coupling. Yet, consensus optimal partitions show that skeleton coupling determines the planted community structure better than any other methods.

In Fig.~\ref{fig6}(B), we explore the performance of the coupling strategies to detect transient communities in the data.  Interestingly, when combined with the Infomap method, diagonal coupling approaches perform very poorly, failing to carry over planted transient communities in each layer. Whereas, the use of these coupling techniques results in a tendency for the community labels of singleton communities to be carried over.  On the other hand, we again find that non-diagonal coupling approaches perform better than diagonal coupling, as indicated by the NMI values (darker shade of red). Indeed, example partitions show that both neighborhood coupling and skeleton coupling can identify the planted \emph{transient communities}. However, neighborhood coupling fails to temporally carry over singletons (similar to Fig.~\ref{fig6}(A)), increasing the total community labels. Similarly, skeleton coupling  performs temporal carry overs of singleton communities more successfully and shows a high correlation between the detected community evolution and ground truth. Moreover, these results are robust across different networks with the same planted dynamic community structure for optimal regions as shown by the low standard deviation for these parameter choices. Lastly, all of these results hold for similar community evolution scenarios which we present in Fig. \ref{supfig1}.

\section{Discussion}
 
Real-world complex systems exhibit dynamical behavior in which the state of the system changes over time. Subsequently, temporal networks and dynamic community detection (DCD) can be used to assess the evolution of network communities.  Here, we examined 5 different DCD methods and showed that current implementations for these methods fail in data with many singleton communities and transient events.  We also found that methods that employed interlayer edge coupling strategies (MMM and Infomap) performed better at identifying singleton communities. This observation led us to focus on DCD methods that utilized interlayer coupling.  However, it is worth noting that the other three algorithms that did not involve interlayer links used rules to determine how community labels were propagated across layers.  It is possible that one could define different rules that might also improve the detection of singleton and transient communities, and other work could pursue this idea.

When utilizing interlayer links, we further found that the use of non-diagonal coupling algorithms additionally resulted in better temporal carryover of community assignments.  We therefore developed a novel non-diagonal interlayer coupling scheme that we call \textit{skeleton coupling}, which incorporates the temporal neighborhood history encoded in the adjacent previous network states in order to algorithmically determine the placement of interlayer edges.  Skeleton coupling outperformed existing interlayer coupling schemes by temporally carrying over both singleton and large community labels in synthetically generated data.

Skeleton coupling builds upon the idea that singleton communities are topologically different than larger size communities. We think of dynamic communities as 1-skeletons (or cliques of partitions) independent of their connectivity in the network. In other words, given a partition of a network into communities, the 1-skeleton of a community is the fully connected subnetwork nodes which doesn't have any other outside connections, discretizing the communities from the rest of the network. Since a singleton community i.e., a 0-skeleton, does not have any edges within, it cannot have a 1-skeleton, whereas a larger size community containing edges between the members of the community can have at least 1-skeleton. Therefore, a 0- and 1-skeleton are topologically different, and they have to be coupled differently. By considering the time evolution of skeletons of communities on a temporal network, skeleton coupling algorithmically links connected components of temporal networks, which corresponds to assigning interlayer edges in the discretized versions of communities in the skeleton representation. 

The development of novel non-diagonal coupling schemes is also motivated by the fact that when diagonal coupling schemes are used, the importance of selecting the proper edge weight, threshold, and DCD method can be additionally complicated. In our parameter space plots of Figs.~\ref{fig5} and \ref{fig6}, we can make a general observation that the value of the interlayer edge weight seems less important than the value of threshold parameter, as seen by the similar color value of the NMI that extends vertically throughout the plots.  In fact, when comparing Figs.~\ref{fig5}A and B using the MMM method, it is clear that the optimal threshold parameter is highly dependent on the type of community event for the diagonal coupling schemes.  This effect is much less pronounced for the non-diagonal coupling schemes.  We again see a similar effect when looking at the performance of Infomap in Fig.~\ref{fig6}.  Here, for diagonal coupling schemes, we again see differences in the regions of optimal method hyperparameters between panels A and B.  Interestingly, there is also more dependence on the choice of edge weight for diagonal coupling schemes used with Infomap.  Further, when employing non-diagonal coupling schemes (neighborhood and skeleton), the optimal regions in Figs.~\ref{fig5} and \ref{fig6} exhibit a much stronger NMI which extends along the entire parameter space (vertically) for both expanding and transient community events, and this observation is independent of the choice of DCD method. This finding suggests that non-diagonal linking schemes such as skeleton coupling can be used as a dimensionality reduction technique since the choice of optimal method hyperparameters is less dependent on the interlayer coupling edge weight, $\omega$, type of community event, and choice of a method. Lastly, note that the standard deviations are low for the optimal region parameter values in all plots in Figs. \ref{fig5} and \ref{fig6}, therefore, this shows the robustness of our results across different stochastic runs of network generation.

While non-diagonal coupling schemes show many advantages, one drawback of skeleton coupling is the computational cost of the given framework. In order to take advantage of skeleton coupling, one needs to apply static community detection to individual snapshots, as skeleton coupling utilizes the static community information in order to determine interlayer edges. This means that the static version of the DCD method needs to run multiple times before running the DCD method on the full temporal network, which clearly increases the computational complexity. However, on short-stacked temporal networks (low number of snapshots), the time consumption of the method is not problematic given that the accuracy of the DCD method is drastically improved.

It is noteworthy that different community detection methods (both static and dynamic) are not deterministic in the sense that they may yield different partitions on each run of the method, although the results may be structurally similar. Although we ran our dynamic community detection methods multiple times on each synthetic neuronal time-series and showed a dynamic consensus partition in our results throughout the text, skeleton links are determined by the same static community detection method for only one single run of the static version of the same method applied at each layer.  in our case, the reason for performing only a single run of the algorithm at each layer was purely computational.  Performing a consensus over the detection at each layer introduces a level of computational complexity that when combined with the large parameter space we were exploring was unreasonable.  In situations where computational complexity was not a concern, one could add an additional step in finding skeleton links by computing an ensemble of static partitions obtained from different algorithmic runs of the static community detection method for each layer $t$, $\{P_{t}\}_{k=0}^{10}$, and then finding the consensus partition of each layer, $CP_{t}$. In this case, one would determine the skeleton links based on this consensus partition.

Finally, we note that the same idea of skeleton coupling can be extended further to higher-order skeletons. For example, a community of size 2 is also topologically different than a community of size 3 and more, as the size 2 community can at most have a 1-skeleton, whereas the larger community can have at least a 2-skeleton which corresponds to the filled in triangles in the corresponding simplicial complex. In general, community sizes and dimensions of the associated skeletons are correlated and a community of size $k$ can have at most a $(k-1)$-skeleton, which distinguishes it from larger size communities. Within our presented framework, additionally utilizing higher-order skeletons to select interlayer edges would result in the addition of subcases of the \textit{`case iv'} section of the presented algorithm. We anticipate that incorporating higher-order skeletons would improve the performance of DCD methods by introducing greater differences in topological coupling between layers. However, here, we only focus on skeletons up to dimension 1 due to the previously discussed computational concerns. 

In summary, this work fills an important gap in the literature by comparing and contrasting various DCD methods and their optimal method hyperparameters for performing community detection in temporal networks. In real-world data where the ground truth community structure and evolution is not known, understanding how network construction and choice of DCD method affects the outcome of the detected community evolution is essential.  In data sets with expected singleton and transient communities, we therefore recommend the use of non-diagonal coupling algorithms such as skeleton coupling to improve method performance and provide more accurate representation of community evolution.  This is especially important in functional neuronal networks where we expect many independent (uncorrelated) neurons due to an under sampling of the population which translates to the presence of many singleton communities, or the rapid evolution of cell assembly composition which leads to many transient and evolving communities. Finally, although our work was motivated from networks derived from neuroscience data, the methodology presented in the paper here is not limited to such data, and the same recommendations hold for any temporal network in which one might suspect singleton and/or transient communities in the data.

\section{Methods}\label{sec:methods}

\subsection{Code and Data availability}
Code for the implementation of skeleton coupling can be found at \cite{temporal_networks} and an accompanying  documentation for this codebase can be found at \cite{temporal_networks_doc}.

\subsection{DCD methods}
In order to account for stochasticity within the DCD algorithms, each following DCD method was run 10 times in the presented results. A consensus partition was then calculated according to the maximum overlap consensus (MOC) method \cite{consensusdissesnsus} across all the algorithmic runs.  Please note that all partitions represent the community assignment of each node in each layer and that a given community label can exist across layers (over time) and that communities can be born and die in temporal network data.

\subsubsection{Multilayer modularity maximization (MMM)}
Modularity assesses partition quality based on a comparison between the connectivity of nodes within a community and between communities, relative to what would be expected in a null model \cite{MMM}. 
In our case, communities within layers are compared to the configuration model \cite{configuration} by utilizing the Leiden solver \cite{leidenalg} (instead of commonly used Louvain algorithm \cite{louvainalg}). In Fig.\ref{fig2}, We explored method performance as a function of the edge threshold and resolution parameter, $(T, \gamma)$. For this analysis, all calculations were performed assuming uniform diagonal coupling with interlayer edge weight $\omega = 0.1$.  For the analysis in Fig.\ref{fig5}, we selected the optimal interlayer edge weight found from the analysis done in Fig.\ref{fig2} (Fig.\ref{fig5}A, $\gamma = 0.94$; Fig.\ref{fig5}B, $\gamma = 1.46$) and fixed this parameter in order to explore the $(T, \omega)$ parameter space for different interlayer coupling configurations.

\subsubsection{Infomap}
Infomap determines community structure based on the visiting times of nodes by random walkers via the map equation \cite{mapeq,multimapeq}. We used the python API \cite{infomap} with a directed flow model and optimized a two-level partition in order to run our analyses.  We explored method performance as a function of the edge threshold and multilayer relax rate,  $(T, \rho)$. For this analysis, all calculations were performed assuming uniform diagonal coupling with interlayer edge weight $\omega = 0.1$.  For the analysis in Fig.\ref{fig5}, we selected the optimal multilayer relax rate found from the analysis done in Fig.\ref{fig2} (Fig.\ref{fig5}A and B, $\rho = 0.2$) and fixed this parameter in order to explore the $(T, \omega)$ parameter space for different interlayer coupling configurations.

\subsubsection{Dynamic stochastic block model (DSBM)} 
Stochastic block models determine community structure by trying to fit generative models to known properties of the data \cite{dsbm,dsbm2}. In our experiments, 
we utilized  \emph{LayeredBlockState} in the graph-tool API \cite{graphtool} with overlapping model and edge covariates chosen as `real-exponential' \cite{dsbm}. We ran our analyses with and without degree correction $\Delta$ ($1$ and $0$, respectively), which we included in $(T,\Delta)$ parameter spaces as two different rows in Fig.\ref{fig2}.

\subsubsection{Dynamic plex propagation method (DPPM)}
Dynamic plex propagation method is a generalization of the clique percolation method (CPM) \cite{cpm}. DPPM relaxes the condition on the definitions of communities, which were $n$-cliques in CPM, into $k$-plexes on $n$ nodes \cite{dppm}. The method runs on individual layers of the network to find the topologically clustered plexes used to define static communities.  These community labels are then separately carried over across snapshots for mapping and matching. We used $n = k+2$ in our analyses for $k=2,3$ as indicated in the rows of the $(T,k)$ parameter space shown in Fig.\ref{fig2}.  We additionally note that one major drawback of the method is its computational complexity which limited the use of this method in our experiments.

\subsubsection{Tensor Factorization}
Tensor factorization determines community structure by approximating the bases of a vector space corresponding to the temporal network. The method takes the desired number of communities, $\eta$ as input, however, this quantity is usually not known in real-world data \cite{tensorfact}.  We therefore explore the $(T,\eta)$ parameter spaces in Fig.\ref{fig2}.  We used a random initialization of the factorization of the tensors with 500 iterations. We averaged the first two factors (x,y-dimensions of the tensors) and multiplied it by the third dimension (time axis) of the basis elements in order to obtain community labels.

\subsection{Synthetic Data Generation}
We simulated neuronal activity of a population of $N=78$ synthetic neurons using a homogeneous Poisson process with firing rates are chosen uniformly at random between $4$ and $40$ for each community. In order to account for stochasticity in our experiments, we studied $k=10$ independent networks. Similarity between firing patterns of neurons was assessed using the pairwise maximum cross-correlation between neurons, meaning that the goal of performing DCD on this data set was to detect groups of neurons with similar firing patterns.  To create a temporal network from the time-series data in our experiments, we used window-size of $\tau=1000$ms and $t_{max}=6$ to create snapshots (layers) of the network.

In order to simulate monotonically growing communities, a master spike train of length $t_{max}\times \tau$ was generated with a randomly selected spike rate and jittered $\pm 5$ms to create the master community. The size of the master community was randomly selected from different distributions (uniform, gaussian or exponential) in order to ensure robust results. Next, at every $1000$ms, we generated independent spike trains of a given size (drawn from the same distribution as the master) that synchronized their spiking activity with the master community (again by jittering the master spike train). This process lead to singleton communities joining the master community at every time window.

For non-monotonic events, we input number of communities $N_{c}$ we desire at each snapshot into our time-series generating pipeline. We chose $N_{c}=3$ for each snapshot in the results presented in Figs.\ref{fig5} and \ref{fig6}.  Within each layer, we created master spike trains that were jittered to generate the associated community. Then we `spaced' these communities by generating independently firing spike trains. As a result, in each layer, we had $N$ neurons where some were distributed into communities and some were singletons. If communities in adjacent layers intersected more than $33\%$
, we assigned them the same dynamic community label for determining their planted community labels.

After generating time series for both monotonic and non-monotonic community events, we divided the time series into windows of length $\tau=1000$. We computed the pairwise maximum cross-correlation in each window to build snapshot representations of temporal networks that represent the underlying planted community structure. Finally, before applying any DCD methods, we padded the first and the last layers of every snapshot by the first and last static snapshot to avoid end point issues.  Note that the community partitions in the padded layers was discarded after the method was run.

\subsection{Interlayer coupling}
\subsubsection{Uniform diagonal coupling}
Uniform diagonal coupling was performed by adding undirected interlayer links between a node and itself in adjacent snapshots i.e., for every node $i^{t} \in V$ where $t \in \{1,2,..,t_{max}-1\}$, we assign an interlayer edge of constant weight $\omega$, $(i^{t},i^{t+1})$.

\subsubsection{Diagonal coupling with local updates}
Diagonal coupling with local updates is topologically the same technique as diagonal coupling, but in this case, the interlayer edge weights $\omega_{i}^{t}$ are allowed to vary. Given a constant edge weight $\omega$, $\omega_{i}^{t}$ is equal to $\omega.s$ if the change between nodal attribute at layer $\mathsf{G}^{t}$ and $\mathsf{G}^{t+1}$ is less than (or equal to) $y$ standard deviations, and is equal to $\omega$ if it is greater than $y$ standard deviations as described in \cite{localupd}. We use firing rate as our nodal attribute and take $s = \frac{1}{100}$ and $y = 0.5$ in our experiments.

\subsubsection{Neighborhood coupling}
For neighborhood coupling, we connect $i^{t}$ and $i^{t+1}$ with a uniform edge weight $\omega$. Next, we determine a neighborhood of a node $i^{t}$ where $t\in \{1,2,..,t_{max}-1\}$ in an adjacent layer based on the strength of intralayer edges of $i^{t}$. We sort all the neighbors of $i^{t}$, $\mathsf{N}_{i}^{t}$, in descending order of connection strength and take only the first $p\%$ of these neighbors $\{j^{t+1}\}_{j\in \mathsf{N}_{i}^{t}}$ as the set of maximal neighbors which we couple with $i^{t}$ where $p=10$. We determine the edge weights according to a similar protocol as described in \cite{neigborhoodcpl} for normalizing the weights $\omega$ of intralayer outlinks and using Jensen–Shannon divergence. In particular, the interlayer edge weight between $i^{t}$ and $j^{t+1}$, $\bar{\omega}_{i^{t},j^{t+1}}$, is:
$$\bar{\omega}_{i^{t},j^{t+1}} = 1- JSD(\dfrac{\mathsf{G}^{t}_{(i,j)}}{\sum_{j,j\neq i}\mathsf{G}^{t}_{(i,j)}}.\omega)^{2}$$

\subsection{Optimal regions}
We determine the optimal regions of method parameters by taking the argmax of the  maximal normalized mutual information (NMI) within the parameter planes, $(T,\cdot)$.  The corresponding \emph{example partitions} are then shown for this parameter set. Although the same maximum value can occur at multiple $(T,\cdot)$ pairs, we only display one example partition since different maximal example partitions generally do not differ structurally.  In Figs.\ref{fig5} and \ref{fig6}, we choose the non-optimal regions by keeping the interlayer edge weight, $\omega$, the same and varying the intralayer edge threshold, $T$.

\subsection{Evaluating partition quality}
We compare the performance of dynamic community detection methods with respect to a ground truth which we consider to be our planted community labels. Note that because different DCD methods use different definitions of the optimal community, we expect that different methods will detect different community partitions.  Additionally, when applying DCD methods to real-world data, there is no known ground truth for comparison \cite{groundtruth} which is why we explore synthetic data sets with different planted community evolution.  Importantly, however, when exploring different interlayer coupling strategies, we are making comparisons about method performance within a single DCD method, meaning that we are measuring quantifiable insights about differences in method performance as a function of interlayer coupling independently of the quality function used by the method. We measure the similarity between true community labels $U$ and predicted community labels $V$ by calculating the \emph{normalized mutual information} (NMI) \cite{nmi1,nmi2}:
\begin{equation}\label{eq1}
	NMI(U,V) = \dfrac{I(U,V)}{max(H(U),H(V))}
\end{equation}
where $H(\cdot)$ is the entropy and $I(\cdot,\cdot)$ is the mutual information which we choose to normalize by the maximum entropy of the labels since this approach works better with overlapping communities \cite{nmimax}. Also note that the community labels $U$ and $V$ are for node-layer elements (i.e., each node has a community label that represents its community assignment in each layer). In addition, we illustrate the results in which we measure the quality of the partition by other metrics (ARI, NVI and Jaccard index) in Appendix C `Partition quality metrics'.

\section*{Acknowledgment}

This work was supported by the National Science Foundation (SMA-1734795 to S.F.M.).

\bibliographystyle{unsrt}
\bibliography{main.bib}
\newpage

\appendix
\renewcommand\thefigure{\thesection.\arabic{figure}}    
\setcounter{figure}{0}   

\section{Skeleton coupling algorithm}
We give the pseudocode for our skeleton coupling algorithm in this section. Note that the provided algorithm is not an optimized version, but  the time complexity is at most $\mathsf{O}(N^{2}.t_{max})$. 

Let $\mathsf{P}_{t} = \{C_{1}^{t},C_{2}^{t},...,C_{p}^{t}\}$ and $\mathsf{P}_{t+1} = \{C_{1}^{t+1},C_{2}^{t+1},...,C_{r}^{t+1}\}$ be the partitions for snapshots $\mathsf{G}^{t}$ and $\mathsf{G}^{t+1}$, respectively, obtained by a static community detection method for a node-aligned temporal network $\mathsf{T}=\{\mathsf{G}^{1},\mathsf{G}^{2},...,\mathsf{G}^{t_{max}}\}$. The interlayer links from a node $i^{t}$ to the nodes in snapshot $\mathsf{G}^{t+1}$ can be obtained by the output of the `Algorithm 1'; the dictionary $SC[i,t]$.

\begin{algorithm}[H]
	\caption{Skeleton coupling}
	\bindent
	\begin{algorithmic}[1]
		\FOR{$t = 1$ \TO $t_{max}-1$} 
			\FOR{$\alpha = 1$ \TO $p$}
				\FOR{$i$ in $C_{\alpha}^{t}$}
					\STATE{Find $\beta$ such that $i \in C_{\beta}^{t+1}$}
					\IF{$|C_{\alpha}^{t}|>1$}
						\IF{$|C_{\beta}^{t+1}|>1$}
							\STATE{$SC[i,t] \longleftarrow{[C_{\beta}^{t+1}]}$ }  \hfill \COMMENT{case iv}
						\ELSE
							\STATE{$SC[i,t] \longleftarrow{[]}$}      \hfill  \COMMENT{case iii}
						\ENDIF
					\ELSE
						\IF{$|C_{\beta}^{t+1}|>1$}
							\STATE{$SC[i,t] \longleftarrow{[]}$ } \hfill \COMMENT{case ii}
						\ELSE
							\STATE{$SC[i,t] \longleftarrow{[i^{t+1}]}$} \hfill \COMMENT{case i}
						\ENDIF
					\ENDIF
				\ENDFOR
			\ENDFOR
		\ENDFOR
		
	\end{algorithmic}
	\eindent
\end{algorithm}

\section{Further community evolution scenarios}

\setcounter{figure}{0}   

In the main text, we studied two types of community evolution scenarios (i.e., monotonic and non-monotonic events) in order to compare the performance of our proposed skeleton coupling algorithm combined with MMM and Infomap methods in Figs.\ref{fig5} and \ref{fig6}, respectively. In this section, we provide additional similar experiments on time series with a planted dynamic community structure which represents monotonic and non-monotonic events.

\begin{figure}[!h]
	\centering
	\includegraphics[width = 1\linewidth]{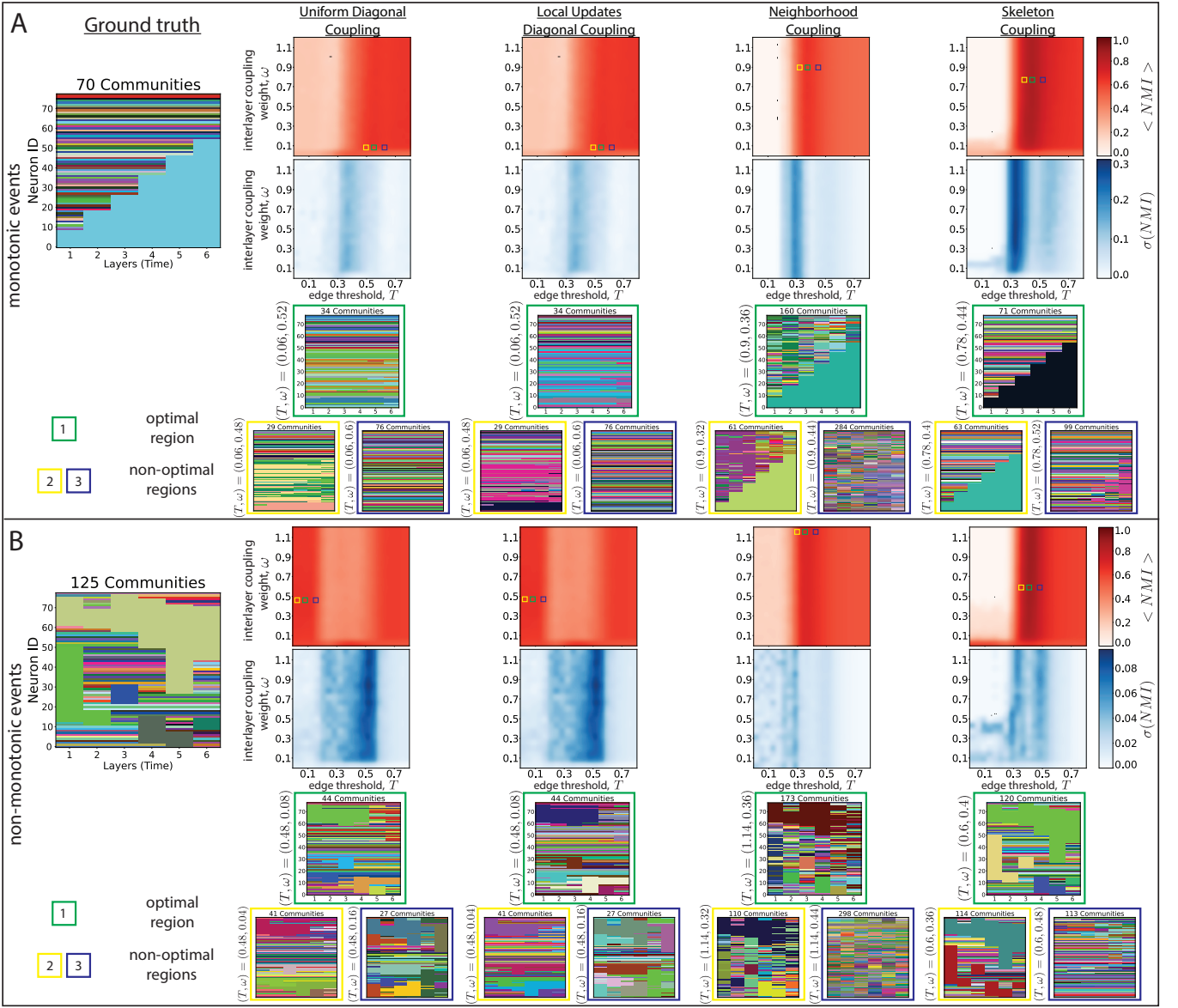}
	\caption{
		\textbf{Comparison of different interlayer coupling strategies for MMM for additional community evolution scenarios.}
		Comparison of different interlayer coupling strategies for $N=78$ neurons undergoing \textbf{A.} expanding community events with 70 total community labels and \textbf{B.} transient community events with a total of 125 community labels. Community labels are depicted by color. Parameter landscapes show method performance measured by the average NMI ($<NMI>$) and standard deviations ($\sigma(NMI)$) in the $(T,\omega)$ parameter space across different stochastic runs.  The configuration model was chosen as the null model and the resolution parameter was equal to $0.98$ for the expanding events in A and $1.3$ for transient events in B. Under each parameter space, we illustrate \emph{example consensus partitions} found by the MMM method within the bounds of optimal and non-optimal regions highlighted by green, yellow and blue rectangles, in the order of descending NMI, respectively.
	}
	\label{supfig0}
\end{figure}

First one of these experiments is again a monotonic growth scenario in which an initial master community recruits groups of singleton communities at each time window as described in Section `Simulation of neuronal data'  and Methods section `Synthetic data generation'. Differently this time, the master growing community does not recruit all the neurons at the end of the time-series. See the ground truths in panel A of Figs. \ref{supfig0} and \ref{supfig1}. 

Second one of these experiments is exactly the same as the non-monotonic events we described in the Methods section `Synthetic data generation'. However, this time we choose $N_{c}=2$ in panels B of Figs. \ref{supfig0} and \ref{supfig1}. By doing so, we achieve a different community size vs. network size ratio by keeping the number of layers and total number of neurons the same for both of these scenarios.

\begin{figure}[!h]
	\centering
	\includegraphics[width = 1\linewidth]{new_heuristics_infomap_non_optimal_2.pdf}
	\caption{
		\textbf{Comparisons of different interlayer coupling strategies for Infomap  for additional community evolution scenarios.}
		Comparison of different interlayer coupling heuristics for $N=78$ neurons undergoing \textbf{A.} expanding community events with 70 total community labels and \textbf{B.} transient community events with a total of 125 community labels. Community labels are depicted by color. Parameter landscapes show method performance measured by the average NMI ($<NMI>$) and standard deviations ($\sigma(NMI)$) in the $(T,\omega)$ parameter space across different stochastic runs. The multilayer relax rate was set to $0.2$ in both panels. Under each parameter space, we illustrate \emph{example consensus partitions} found by the Infomap method within the bounds of optimal and non-optimal regions highlighted by green, yellow and blue rectangles, in the order of descending NMI, respectively.
	}
	\label{supfig1}
\end{figure}

\section{Partition quality metrics}
\setcounter{figure}{0}   

In the main text, we used normalized mutual information (NMI) as our main metric to study partitions with respect to a ground truth. We test the robustness of our results for skeleton coupling by studying the parameter spaces which we examined in Figs.\ref{fig5} and \ref{fig6} via adjusted rand index (ARI), normalized variation information (NVI) and Jaccard index in Figs.\ref{supfig2} and \ref{supfig3}, respectively. Our results show that the optimal and non-optimal regions with respect to these partition quality metrics coincide with the ones obtained from NMI. Hence, the results presented in the main text remain to be robust.

\begin{figure}[!h]
	\centering
	\includegraphics[width = 1\linewidth]{new_heuristics_mmm_metrics.pdf}
	\caption{
		\textbf{Comparisons of parameter landscapes for different information measures and interlayer coupling strategies with MMM.}
		We study the resulting partitions by comparing them with a ground truth as a function of ARI, NVI and Jaccard index.
		\textbf{A.} Monotonic events from Fig.\ref{fig5}.
		\textbf{B.} Non-monotonic events from Fig.\ref{fig5}.
		\textbf{C.} Monotonic events from Fig.\ref{supfig0}.
		\textbf{D.} Non-monotonic events from Fig.\ref{supfig0}.
	}
	\label{supfig2}
\end{figure}

\begin{figure}[!h]
	\centering
	\includegraphics[width = 1\linewidth]{new_heuristics_infomap_metrics.pdf}
	\caption{
		\textbf{Comparisons of parameter landscapes for different information measures and interlayer coupling strategies with Infomap.}
		We study the resulting partitions by comparing them with a ground truth as a function of ARI, NVI and Jaccard index.
		\textbf{A.} Monotonic events from Fig.\ref{fig6}.
		\textbf{B.} Non-monotonic events from Fig.\ref{fig6}.
		\textbf{C.} Monotonic events from Fig.\ref{supfig1}.
		\textbf{D.} Non-monotonic events from Fig.\ref{supfig1}.
	}
	\label{supfig3}
\end{figure}

\section{Maximum overlapping distances between stochastic runs}

\setcounter{figure}{0}   

\begin{figure}[!h]
	\centering
	\includegraphics[width = 1\linewidth]{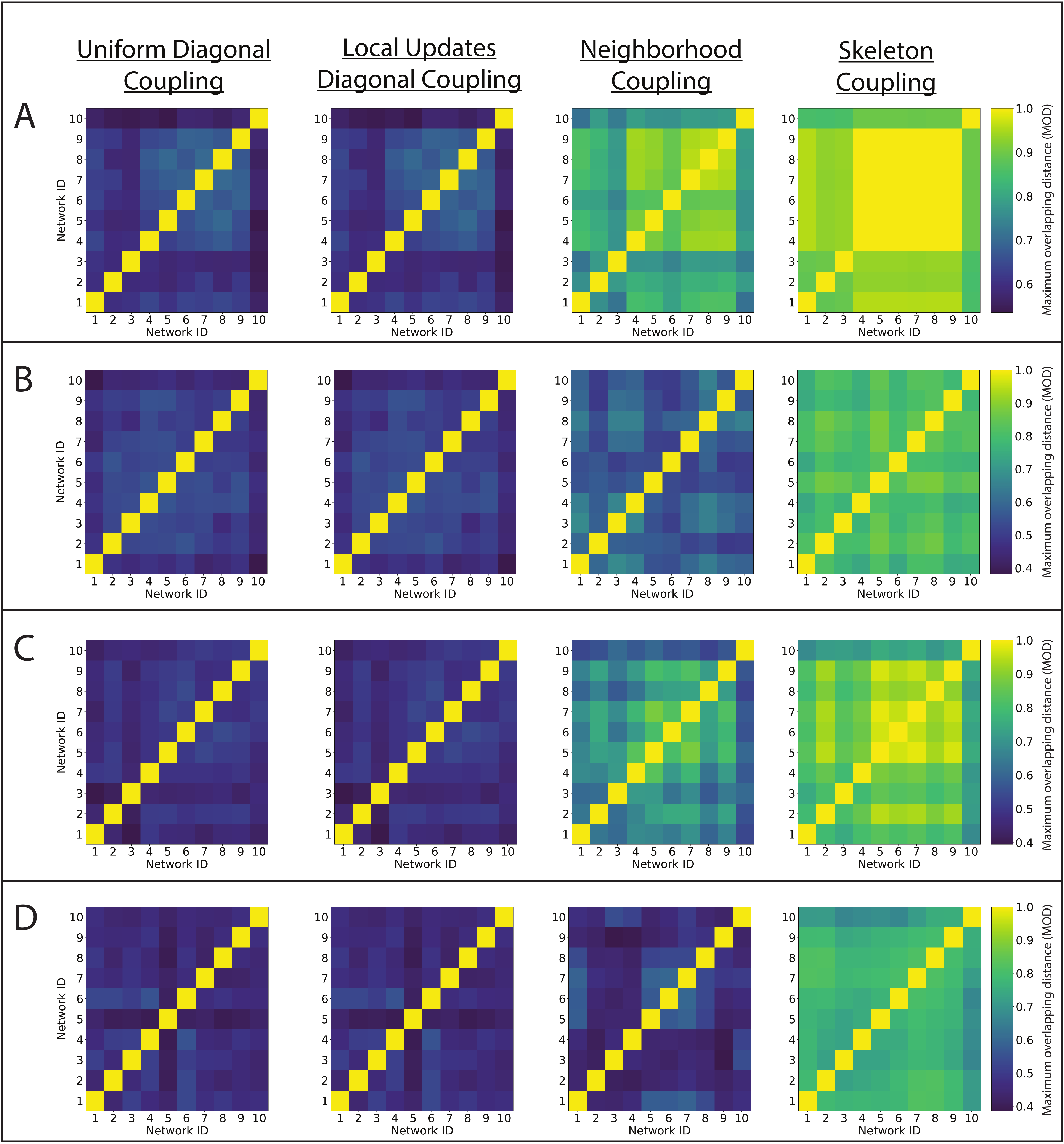}
	\caption{
		\textbf{Maximum overlapping distances between the optimal partitions of MMM in each stochastic run.}
		MOD between the optimal partitions across all stochastic runs in 
		\textbf{A.} Monotonic events from Fig.\ref{fig5}.
		\textbf{B.} Non-monotonic events from Fig.\ref{fig5}.
		\textbf{C.} Monotonic events from Fig.\ref{supfig0}.
		\textbf{D.} Non-monotonic events from Fig.\ref{supfig0}.
	}
	\label{supfig4}
\end{figure}

\begin{figure}[!h]
	\centering
	\includegraphics[width = 1\linewidth]{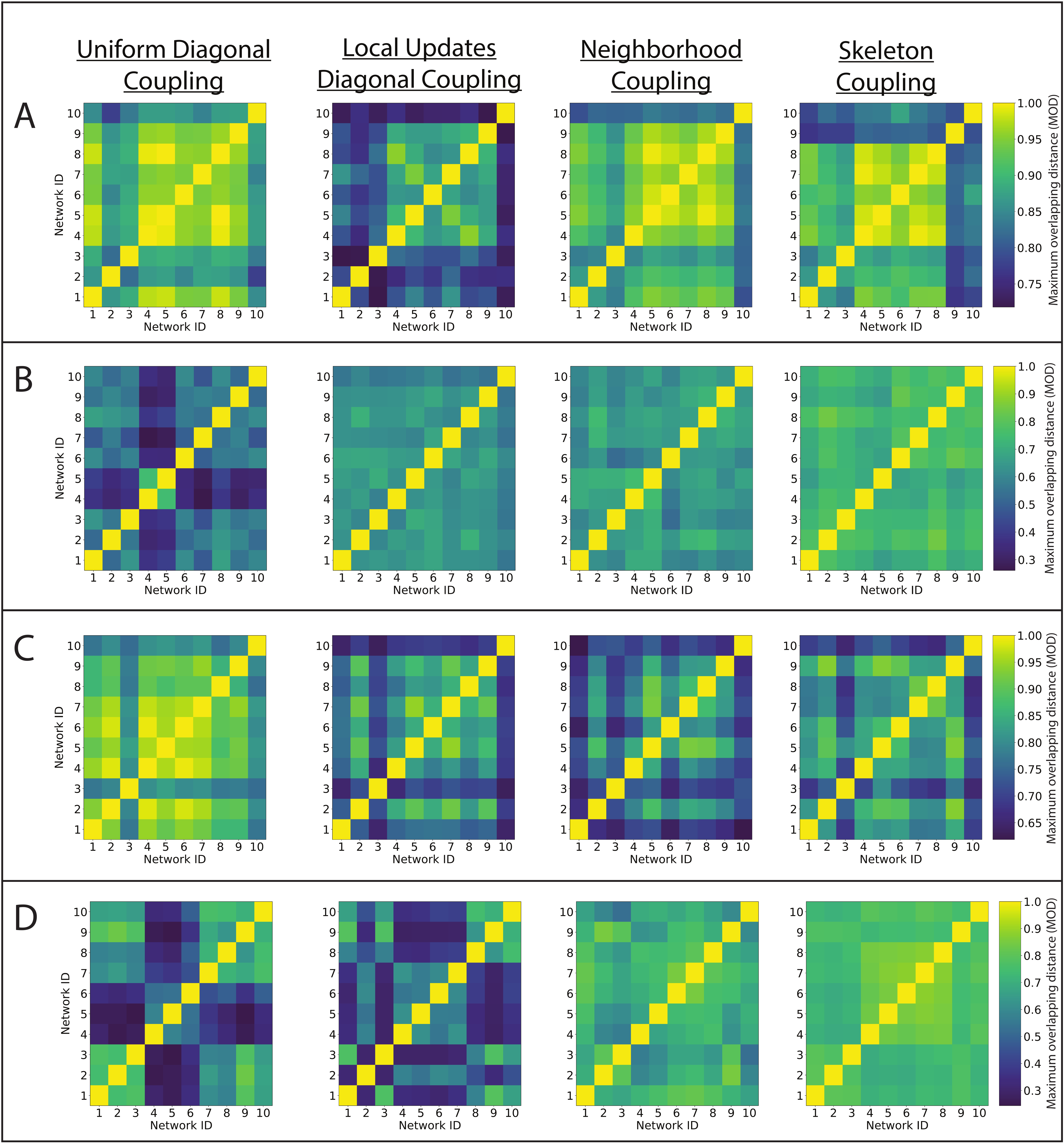}
	\caption{
		\textbf{Maximum overlapping distances between the optimal partitions of Infomap in each stochastic run.}
		MOD between the optimal partitions across all stochastic runs in 
		\textbf{A.} Monotonic events from Fig.\ref{fig6}.
		\textbf{B.} Non-monotonic events from Fig.\ref{fig6}.
		\textbf{C.} Monotonic events from Fig.\ref{supfig1}.
		\textbf{D.} Non-monotonic events from Fig.\ref{supfig1}.
	}
	\label{supfig5}
\end{figure}

As we mentioned in the text, we determine the optimal and non-optimal regions according to the consensus algorithm \cite{consensusdissesnsus} across different stochastic runs. In Figs. \ref{supfig4} and \ref{supfig5}, we study the maximum overlapping distance (MOD) between the optimal partitions of each stochastic generation of a network \cite{consensusdissesnsus}. The higher the MOD is the more similar two partitions are to each other. Each network correspond to a different time series with different spike rates such that the planted dynamic community structure is the same.

In Fig. \ref{supfig4}, we show MOD distances between the optimal partitions for each independently generated time-series with the same community structure obtained from MMM method. From top to bottom panels, we focus on different community evolution scenarios studied in Fig.\ref{fig5}A), Fig.\ref{fig5}B), Fig.\ref{supfig0}A) and Fig.\ref{supfig0}B), respectively. One can observe that, MOD distances are quite low for diagonal coupling approaches, moderate on average for neighborhood coupling and high for skeleton coupling in each scenario. These findings suggests that MMM combined with skeleton coupling yields robust results independent of the stochasticity in the benchmark networks.

Next, in Fig. \ref{supfig5}, we show the exact same results, but for Infomap this time. Similarly, from top to bottom panels, we focus on different community evolution scenarios studied in Fig.\ref{fig6}A), Fig.\ref{fig6}B), Fig.\ref{supfig1}A) and Fig.\ref{supfig1}B), respectively. Although it is hard to compare different heuristics and make a deduction about the resulting MOD distances between optimal partitions in this case, we note that MOD distances for Infomap combined with skeleton coupling are not low, which still suggests relatively robust results.

\end{document}